# THE GLOBULAR CLUSTERS OF DWARF SPHEROIDAL GALAXIES

## SIDNEY VAN DEN BERGH

Dominion Astrophysical Observatory, Herzberg Institute of Astrophysics,
National Research Council of Canada, 5071 West Saanich Road, Victoria, British Columbia,
V8X 4M6, Canada; vdb@dao.nrc.ca

## ABSTRACT

The possibility that the Sagittarius dwarf spheroidal might have formed as a Searle-Zinn fragment in the outer halo of the Galaxy is discussed. Arguments in favor of this hypothesis are: (1) The luminosity distribution of globular clusters in both Sagittarius and in the outer halo ($R_{gc} > 80$ kpc), appear to be bimodal with peaks near $M_V \sim -5$ and $M_V \sim -10$, and (2) the globular clusters in both Sgr, and in the outer halo, have a significantly larger age spread than do the globulars in the inner halo of the Galaxy. However, a counter argument is that only one of the four globulars associated with Sgr has the large half-light diameter that is diagnostic of outer halo clusters. The absence of globular clusters from all Local Group dwarf spheroidal galaxies fainter than $M_V = -12$ shows that their specific globular cluster frequency must be lower than it is in the Fornax dwarf spheroidal. This result suggests that Fornax may have had an unusual evolutionary history.

Subject headings: galaxies: spheroidal - galaxies: star clusters

# 1.  INTRODUCTION

The fact that dwarf spheroidals are galaxies, rather than over-sized globular clusters, is attested to by the fact that they contain dark matter, and that they are themselves sometimes embedded in small systems of globular clusters. In the Local Group such globular clusters are known to be associated with Fornax (Hodge 1961), and Sagittarius (Ibata, Gilmore & Irwin 1994).  A globular also appears to be located in the dSph galaxy F8D1 in the M 81 group (Caldwell et al. 1998).  No globulars are found in any of the Local Group galaxies that are less luminous than Fornax and Sagittarius.  It is of some interest to ask what constraints this places on the value of the specific globular cluster frequency in low-luminosity dwarf spheroidals.

## 2.  SPECIFIC GLOBULAR CLUSTER FREQUENCY IN
## FAINT DWARF  SPHEROIDALS

The specific globular cluster frequency S (Harris & van den Bergh 1981) is defined as

$$S = N \ 10^{0.4(M_v + 15)} \, , \tag{1}$$

in which N is the total number of globular clusters associated with a galaxy having integrated magnitude $M_V$.  In other words, the specific frequency is the number of globulars per $M_V = -15$ of parent galaxy light.  For 53 early-type galaxies van den Bergh (1998a) found <S> = 5.8, with individual values ranging from S = 21 in UGC 9799, which is the central dominant galaxy in the cluster A 2052, to S = 0 in M 32 (from which some clusters might have been stripped by tidal interactions with M 31).  The specific cluster frequency in Fornax is high (Harris 1991).  However, its exact value is not well determined because the integrated magnitude of this large faint object is difficult to measure.  Demers, Irwin & Kunkel (1994) discuss previous determinations, which range from $M_V$ = -12.3 to $M_V$ = -13.6.  They conclude that Fornax has $M_V$ = -13.1.  With this luminosity S(Fornax) = 29, which is larger than  that for any other early-type galaxy listed in the compilation of van den Bergh  (1998a).  The specific globular cluster frequency for the Sagittarius dwarf spheroidal is impossible to determine because the integrated magnitude of this large and tidally distorted galaxy cannot be measured with any degree of confidence.  From application of the virial theorem to the velocity measurements of individual stars in the Sagittarius system Ibata et al. (1997) derive M(Sgr) ~1.5  x $10^8$ $M_\odot$.  This yields a specific frequency of  ~2.7 globulars per $10^8$ $M_\odot$.  According to Mateo (1998) M(For) = 6.8 x $10^7$ $M_\odot$, so that it has 4.1 globulars  per $10^8$ $M_\odot$.  If one were to *assume* that the Sagittarius and Fornax dwarf spheroidals have similar M/L values, then it would follow that the specific cluster frequency of Sagittarius is less than half of that for Fornax.

        From the compilation of van den Bergh (1994) it is found that the 11 then known Local Group dwarf spheroidals fainter than $M_V$ = -12 have a combined

integrated magnitude $M_V$ = -13.3. None of these dwarf galaxies contains a globular cluster. The total number of clusters that one expects to be associated with a galaxy having this integrated magnitude is 0.21 x S. From Poisson statistics one finds that the *a priori* probability of finding no globular clusters in such a galaxy is 0.81 for S = 1, 0.35 for S = 5, and 0.015 for S = 20. The hypothesis that the faint Local Group galaxies resemble Fornax and have S = 29 can be ruled out at the 99.8% confidence level. Even lower probabilities would be obtained if one were to include galaxies of type dIr/dSph, such as Phoenix and Pisces, and the more recently discovered faint dwarf spheroidals. It is therefore concluded that the specific globular cluster frequency for faint Local Group dwarf spheroidals is probably < 5. The fact that Fornax has a higher S value than any other bright or faint early-type galaxy suggests that it has had an unusual evolutionary history.

### 3. THE SAGITTARIUS DWARF AND ITS COMPANIONS

The globular clusters NGC 6715 (= M 54), Arp 2, Terzan 7 and Terzan 8 appear to be associated with the Sagittarius dwarf spheroidal (Ibata et al. 1994). M 54 has $M_V$ = -9.96, which is only slightly fainter than the brightest Galactic globular ω Centauri, for which $M_V$ = -10.24. The other three clusters are all quite faint and have $M_V$ = -5.24 (Arp 2), $M_V$ = -5.00 (Ter 7), and $M_V$ = -5.06 (Ter 8). A comparison between the luminosity distributions of Galactic globular clusters

(Harris 1997) and that of the Sagittarius globulars is shown in Figure 1.  This

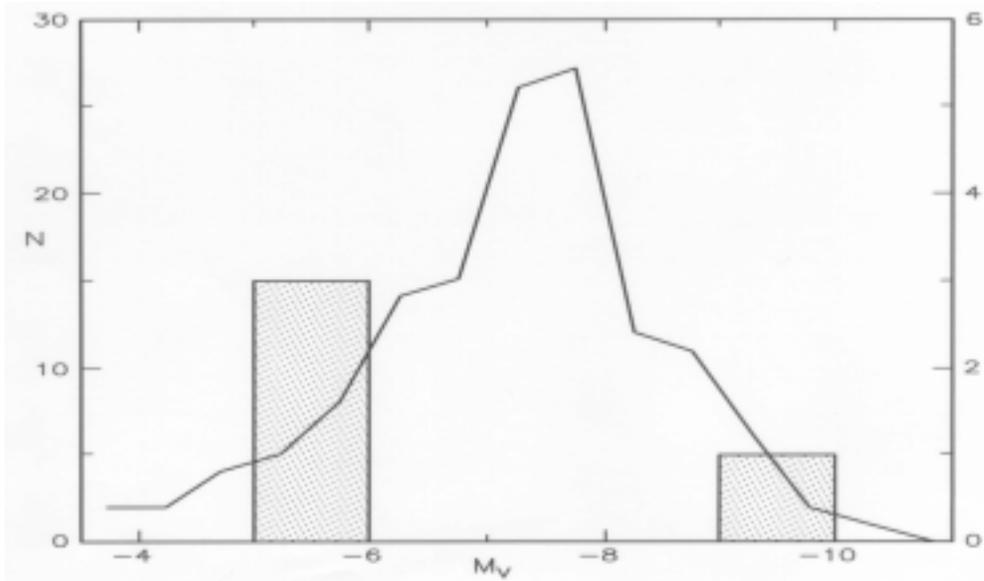

Figure 1

figure suggests that the globular clusters associated with the Sagittarius dwarf

might have a bimodal luminosity distribution.  A Kolmogorov-Smirnov test

shows that there is only a 10% probability that the Galactic and Sagittarius

globular clusters were drawn from the same parent population. The luminosity

distribution of the Sagittarius globulars is reminiscent of that in the outer halo of

the Galaxy (Harris 1996)[1], which is shown in Figure 2.  The figure shows that the

---

[1]  An up-dated version of these data at URL

http://www.physics.mcmaster.ca/Globular.html was actually used.

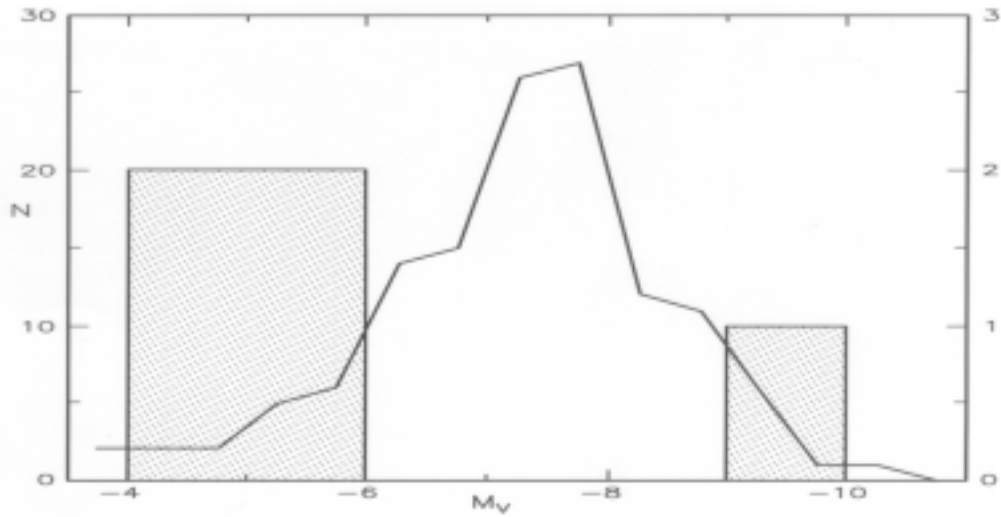

Figure 2

globular clusters in the outer halo ($R_{gc} > 80$ kpc) have a bimodal luminosity function that differs significantly from that of the globulars with $R_{gc} < 80$ kpc. A Kolmogorov-Smirnov test shows that there is only a 4% probability that the inner and outer halo globulars were drawn from the same parent population. It is noted in passing that the outer halo cluster NGC 2419 is the third-brightest Galactic globular cluster.

A second similarity between the globular clusters associated with the Sagittarius dwarf, and those in the outer halo at $R_{gc} > 80$ kpc, is that they have a much larger age range than do the globular clusters in the main body of the Galaxy. According to Montegriffo et al. (1998) the cluster Ter 7 is ~5 Gyr to ~9 Gyr younger than Ter 8. This age range is even larger than that found among the globular clusters in the outer Galactic halo. In view of these similarities it is tempting to speculate that Sagittarius is a Searle-Zinn (1978) fragment formed in the outer halo that has fallen inwards towards the center of the Galaxy. However, a problem with this hypothesis is that the Sagittarius dwarf presently appears to be in a rather short period orbit (Velázquez & White 1995, Ibata et al. 1997, Ibata & Lewis 1998). A possible way out of this dilemma (Zhao 1998) is to assume that the Sagittarius dwarf was scattered into its present orbit during a recent encounter with the Magellanic Clouds. However, such a close encounter between Sgr and LMC + SMC has a low *a priori* probability. A second problem is that only one (Arp 2) of the four globulars associated with Sgr has the large half-light radius that is diagnostic of outer halo clusters (van den Bergh & Morbey 1984). In summary it appears that one must bring in a Scottish "not proven" verdict on the hypothesis that Sgr started its existence as a Searle-Zinn fragment.

Minniti, Meylan & Kissler-Pattig (1996) have argued that the cluster Terzan 7 is too metal-rich to be associated with the Sagittarius dwarf galaxy. However, Marconi et al. (1998) state that "the dominant population [in Sagittarius] is extremely similar to the stellar content of the globular cluster Terzan 7." Therefore it does not seem possible to use metallicity alone as an argument against the hypothesis of Ibata et al. (1994) that Terzan 7 is physically associated with the Sagittarius dwarf. In any case it would be at least as unexpected for an independent Terzan 7, at a Galactocentric distance of ~18 kpc, to be metal-rich!  Velásquez & White (1995) have argued that Terzan 7 might not be associated with Sagittarius because its radial velocity differs from that of NGC 6715, which is often assumed to be located near the dynamical center of the Sagittarius dwarf.  However, the velocity measurements of Da Costa & Armandroff (1995) show that the radial velocity of Arp 2 ( $V_r = 115$ km s$^{-1}$) differs from that of NGC 6715 ($V_r = 142$ km s$^{-1}$) by even more than does that of Terzan 7 ($V_r = 166$ km s$^{-1}$).  Terzan 7 and Arp 2 are separated from NGC 6715 by projected distances of 1.8 kpc and 2.0 kpc, respectively.  They could only have reached such large separations from their parent galaxy if tidal forces resulted in significant velocity differences.  The conclusion that the globular clusters near to Sagittarius have a significant age range would still hold if Terzan 7 turned out not to be physically associated with Sagittarius.  Montegriffo et al. (1998) find that Arp 2 is (4.5 ± 2) Gyr younger than Ter 8, whereas Ter 7 is (7 ± 2) Gyr younger than Ter 8.  Finally Da Costa & Armandroff (1995) have speculated that NGC 6715 = M 54 might be the nucleus of the Sagittarius galaxy.  In a study of spheroidal galaxies in the Virgo cluster (van den Bergh 1986) it was found that the fraction of such objects that are nucleated drops from ~100% at $M_B = -17$ to about 10% near $M_B = -12$.  Furthermore he found that the fraction of nucleated dwarfs was largest in the densest regions of the Virgo cluster.  Both of these observations militate against (but do not exclude) the possibility that M 54 might be the nucleus of the Sagittarius dwarf galaxy.

## 4. CONCLUSIONS

The rather scanty data that are presently available suggest that the luminosity distributions of Galactic globular clusters with $R_{gc} > 80$ kpc, and that of the globulars associated with the Sagittarius dwarf spheroidal, are both bimodal. A second similarity is that the globulars in both Sgr and the outer halo exhibit a larger age range than do the globular clusters with $R_{gc} < 80$ kpc. This suggests that the Sagittarius dwarf may have formed in the outer halo. A difficulty with this hypothesis is that simulations, in fact, appear to indicate that Sgr is on a short- period (~0.7 Gyr) orbit. Possibly this conclusion can be avoided by assuming that the Sagittarius dwarf was scattered into its present short-period orbit by a gravitational interaction with the Magellanic Clouds. However, an argument against the hypothesis that Sgr is a Searle-Zinn fragment is that three of the four globular clusters that are associated with it do not have the large half-light radii that are diagnostic of globular clusters in the outer Galactic halo.

The absence of globular clusters associated with Local Group dwarf spheroidal galaxies fainter than $M_V = -12$ suggests that $S < 5$ for the faintest dwarf spheroidals. This is marginally lower than the average for all early-type galaxies, and significantly lower than S(For). This suggests that the Fornax dwarf spheroidal may have had an unusual evolutionary history.

## FIGURE LEGENDS

Fig. 1.  Comparison between the luminosity distribution of all Galactic globular clusters (curve, scale at left) and of the globular clusters associated with the Sagittarius dwarf (histogram, scale at right).  A Kolmogorov-Smirnov test shows that there is only a 10% probability that these two distributions were drawn from the same parent population.  Note the similarity of the Sagittarius cluster luminosity distribution to that of Galactic halo clusters with $R_{gc} > 80$ kpc, which is shown in Figure 2.

Fig. 2.  Comparison between the luminosity distribution of Galactic globular clusters with $R_{gc} < 80$ kpc (curve, scale at left), and that of the globular clusters with $R_{gc} > 80$ kpc (histogram, scale at right).  A Kolmogorov-Smirnov test shows that there is only a 4% probability that these two distributions were drawn from the same parent population.